\documentclass[aps,prl, twocolumn,amsmath,amssymb,superscriptaddress,floatfix,longbibliography]{revtex4-1}
\usepackage[breaklinks=true,colorlinks,citecolor=blue,linkcolor=blue,urlcolor=blue]{hyperref}

\usepackage[T1]{fontenc}
\usepackage[latin9]{inputenc}
\usepackage{lineno}
\usepackage{textcomp}
\usepackage{amsmath}
\usepackage{amssymb}
\usepackage{bm}
\usepackage{float}
\usepackage{graphicx}
\usepackage{multirow}
 \usepackage{makecell}
 \usepackage{booktabs}
 \usepackage{mathtools}
\usepackage{array}
\usepackage{epsfig,mathrsfs,color,latexsym,subfigure,marginnote}

\newcolumntype{P}[1]{>{\centering\arraybackslash}p{#1}}

\def\be{\begin{equation}}
\def\ee{\end{equation}}

\def\bea{\begin{eqnarray}}
\def\eea{\end{eqnarray}}

\usepackage[colorlinks,linkcolor=blue]{hyperref}

\usepackage{subfigure}
\usepackage{color}

\definecolor{Red}{rgb}{1,0,0}
\definecolor{Blu}{rgb}{0,0,1}
\definecolor{Green}{rgb}{0,1,0}

\makeatother
\usepackage{amssymb}

\usepackage{tikz,xcolor,hyperref}
\definecolor{lime}{HTML}{A6CE39}
\DeclareRobustCommand{\orcidicon}{%
	\begin{tikzpicture}
	\draw[lime, fill=lime] (0,0)
	circle [radius=0.16]
	node[white] {{\fontfamily{qag}\selectfont \tiny ID}};
	\draw[white, fill=white] (-0.0625,0.095)
	circle [radius=0.007];
	\end{tikzpicture}
	\hspace{-2mm}
}

\foreach \x in {A, ..., Z}{%
	\expandafter\xdef\csname orcid\x\endcsname{\noexpand\href{https://orcid.org/\csname orcidauthor\x\endcsname}{\noexpand\orcidicon}}
}

\begin{document}

\title{Switchable large-gap quantum spin Hall state in two-dimensional MSi$_2$Z$_4$ materials class}

\author{Rajibul Islam}
\email{rislam@magtop.ifpan.edu.pl}
\affiliation{International Research Centre MagTop, Institute of Physics, Polish Academy of Sciences, Aleja Lotnik\'ow 32/46, PL-02668 Warsaw, Poland}
\affiliation{Department of Condensed Matter Physics and Materials Science, Tata Institute of Fundamental Research, Colaba, Mumbai 400005, India}

\author{Rahul Verma}
\affiliation{Department of Condensed Matter Physics and Materials Science, Tata Institute of Fundamental Research, Colaba, Mumbai 400005, India}

\author{Barun Ghosh}
\affiliation{Department of Physics, Northeastern University, Boston, Massachusetts 02115, USA}

\author{Zahir Muhammad}
\affiliation{International Research Centre MagTop, Institute of Physics, Polish Academy of Sciences, Aleja Lotnik\'ow 32/46, PL-02668 Warsaw, Poland}
\affiliation{Hefei Innovation Research Institute, School of Microelectronics, Beihang University, Hefei 230013, P. R. China}

\author{Arun Bansil}
\affiliation{Department of Physics, Northeastern University, Boston, Massachusetts 02115, USA}

\author{Carmine Autieri}
\email{autieri@magtop.ifpan.edu.pl}
\affiliation{International Research Centre MagTop, Institute of Physics, Polish Academy of Sciences, Aleja Lotnik\'ow 32/46, PL-02668 Warsaw, Poland}
\affiliation{Consiglio Nazionale delle Ricerche CNR-SPIN, UOS Salerno, I-84084 Fisciano (Salerno), Italy}

\author{Bahadur Singh}
\email{bahadur.singh@tifr.res.in}
\affiliation{Department of Condensed Matter Physics and Materials Science, Tata Institute of Fundamental Research, Colaba, Mumbai 400005, India}

\begin{abstract}
Quantum spin Hall (QSH) insulators exhibit spin-polarized conducting edge states that are topologically protected from backscattering and offer unique opportunities for addressing fundamental science questions and device applications. Finding viable materials that host such topological states, however, remains a challenge. Here by using in-depth first-principles theoretical modeling, we predict large bandgap QSH insulators in recently bottom-up synthesized two-dimensional (2D) MSi$_2$Z$_4$ (M = Mo or W and Z = P or As) materials family with $1T^\prime$ structure. A structural distortion in the $2H$ phase drives a band inversion between the metal (Mo/W) $d$ and $p$ states of P/As to realize spinless Dirac cone states without spin-orbit coupling. When spin-orbit coupling is included, a hybridization gap as large as $\sim 204$ meV opens up at the band crossing points, realizing spin-polarized conducting edge states with nearly quantized spin Hall conductivity. We also show that the inverted band gap is tunable with a vertical electric field which drives a topological phase transition from the QSH to a trivial insulator with Rashba-like edge states. Our study identifies 2D MSi$_2$Z$_4$ materials family with $1T^\prime$ structure as large bandgap, tunable QSH insulators with protected spin-polarized edge states and large spin-Hall conductivity.   
\end{abstract}

\maketitle

{\it Introduction:--}
Following the early studies of two-dimensional (2D) materials~\cite{GrapheneNL,Graphene_Neto,Graphene2005,TMDs2017}, Kane and Mele demonstrated the existence of a quantum spin Hall (QSH) state in graphene in the presence of symmetry-allowed spin-orbit coupling (SOC)~\cite{Kane_Male_QSH_Graphene,PhysRevLett.95.146802}. The QSH state features one-dimensional (1D) conducting helical edge modes inside an insulating bulk due to the nontrivial winding of their electronic states~\cite{RMP_Bansil,RMP_Hasan,Kane_Male_QSH_Graphene,PhysRevLett.95.146802,QSH_SCZhang,Theory_HgTe_QSH,QSH_BS,HgPt2Se3_Th}. These helical edge modes carry symmetry-protected spin-polarized electronic states that hold immense potential to design high-efficiency quantum electronic devices with low dissipation~\cite{LiangFu,PhysRevLett.111.136804,PhysRevLett.100.236601}. The QSH state has been theoretically predicted in a variety of 2D materials and quantum well structures. However, the experimental realization of this state has so far been demonstrated only in HgTe/CdTe and InAs/GaSb quantum wells and thin-films of 1T$^\prime-$WTe$_2$, HgPt$_2$Se$_3$, and Bi$_4$Br$_4$ at ultra-low temperatures~\cite{QSH_HgTe,InAs_GaSb_QSH,WTe2_QSH,HgPt2Se3_exp,Bi4Br4_QSH,Fei2017,Wu18}. A common approach to realize the QSH state is to reduce the thickness of three-dimensional (3D) $Z_2$ topological insulators to drive a 3D to 2D crossover and a band inversion in the surface states. This method has successfully predicted the QSH state in thin films of $Z_2$ topological insulators~\cite{QSH_BS,HgPt2Se3_Th,QSH_Hsin,zhang2021two,PhysRevB.96.041108,Li_2020}.  This approach for realizing the QSH state, however,  remains challenging since it requires complex fine-tuning of either quantum well structure or film thickness to generate an inverted hybridization gap in the surface spectrum~\cite{QSH_BS,HgPt2Se3_Th,QSH_Hsin,zhang2021two,PhysRevB.96.041108,Li_2020,acs.nanolett.5b00648,wang2015tuning, smll.201601207}. In the process, materials properties are often modified, leading to complicated electronic structures and quenching the quantized spin Hall conductance. Finding new strategies for designing 2D materials with large inverted bandgaps to achieve the room temperature QSH state is thus highly desirable. 

Here we present an in-depth first-principles analysis with optimized crystal structures to demonstrate the QSH state in recently introduced 2D materials that are realized via a bottom-up approach without parental analogues~\cite{Science_2D,synthetic_2D,hong2020chemical}. Such synthetic 2D materials provide a new paradigm for engineering designer states with diverse functionalities. Specifically, 2D MoSi$_2$N$_4$ materials were synthesized by passivating high-energy surfaces of non-layered nitrides with Si with remarkable stability under ambient conditions~\cite{hong2020chemical}. They show semiconducting behavior with high carrier mobility, and feature spin-valley locking, gating and thickness-tunable spin polarization, and 2D magnetism and correlation-driven quantum anomalous Hall state, among other properties depending on their compositions~\cite{IslamMoSi2N4,yang2021valley,li2020valley,feng2021valley}. Theoretically predicted properties of these bottom-up synthesized materials are reported to be superior to those of the widely used 2D transition metal dichalcogenides (TMDs)~\cite{hong2020chemical,IslamMoSi2N4,bertolazzi2011stretching,cai2014polarity}. It is not clear, however, if these materials can form polytypic structures and realize the QSH state similar to the 2D TMDs. Here, based on our molecular dynamics simulations and phonon calculations, we predict that the $1T^\prime$ phase of MSi$_2$Z$_4$ (M = Mo or W, and Z = P or As) is stable and realizes a QSH state via a structural distortion from 2H to $1T^\prime$ phase. Our calculated inverted bandgap ($\sim204$ meV) and spin Hall conductivity (SHC) [$\sim1.3 e^2/h$] are found to be higher than the top-to-bottom grown 2D TMDs. We also show that the QSH state of MSi$_2$Z$_4$ can be switched off by driving a topological phase transition via an applied (vertical) electric field. Our work indicates the robust presence of a switchable QSH state in a new polytypic structure of bottom-up grown 2D materials with excellent topological and spintronics properties.  

{\it  Methods:--}
Electronic structure calculations were performed within the framework of density functional theory (DFT) based on the projector augmented wave method using the VASP~\cite{kohan_dft,VASP}. The self-consistent relativistic calculations were performed with a plane wave cut-off energy of 500 eV and a $\Gamma$ centred $6\times 12\times 1$ $k-$mesh for Brillouin zone (BZ) sampling. We used the generalized gradient approximation (GGA) to include exchange-correlation effects~\cite{perdew1996generalized}. The structural parameters were fully optimized until the residual forces on each atom were less than 0.001 eV/{\AA} and the total energy is converged to 10$^{-8}$ eV. For a more accurate treatment of electronic correlations, we also employed the Heyd-Scusseria-Ernzerhof (HSE) hybrid functional with 25\% exact Hartree-Fock exchange~\cite{hse}. Phonon dispersions were calculated with the density functional perturbation theory (DFPT) using the PHONOPY code~\cite{TOGO20151} with a 2$\times$4$\times$1 supercell. The {\it ab initio} molecular dynamics simulations were performed through the Nose-Hoover thermostat at a constant temperature of 300K with a time step of $2fs$~\cite{MD}. We generated material-specific tight-binding Hamiltonians with M $d$, Si $s$ and $p$, and Z $p$ orbitals using the VASP2WANNIER interface~\cite{mostofi2008wannier90} which is ultimately used to elucidate topological properties using the Wanniertools package~\cite{wu2018wanniertools,Greenwanniertools}.

{\it  Results:--}
\begin{figure}[ht!]
\centering
\includegraphics[width=0.49\textwidth]{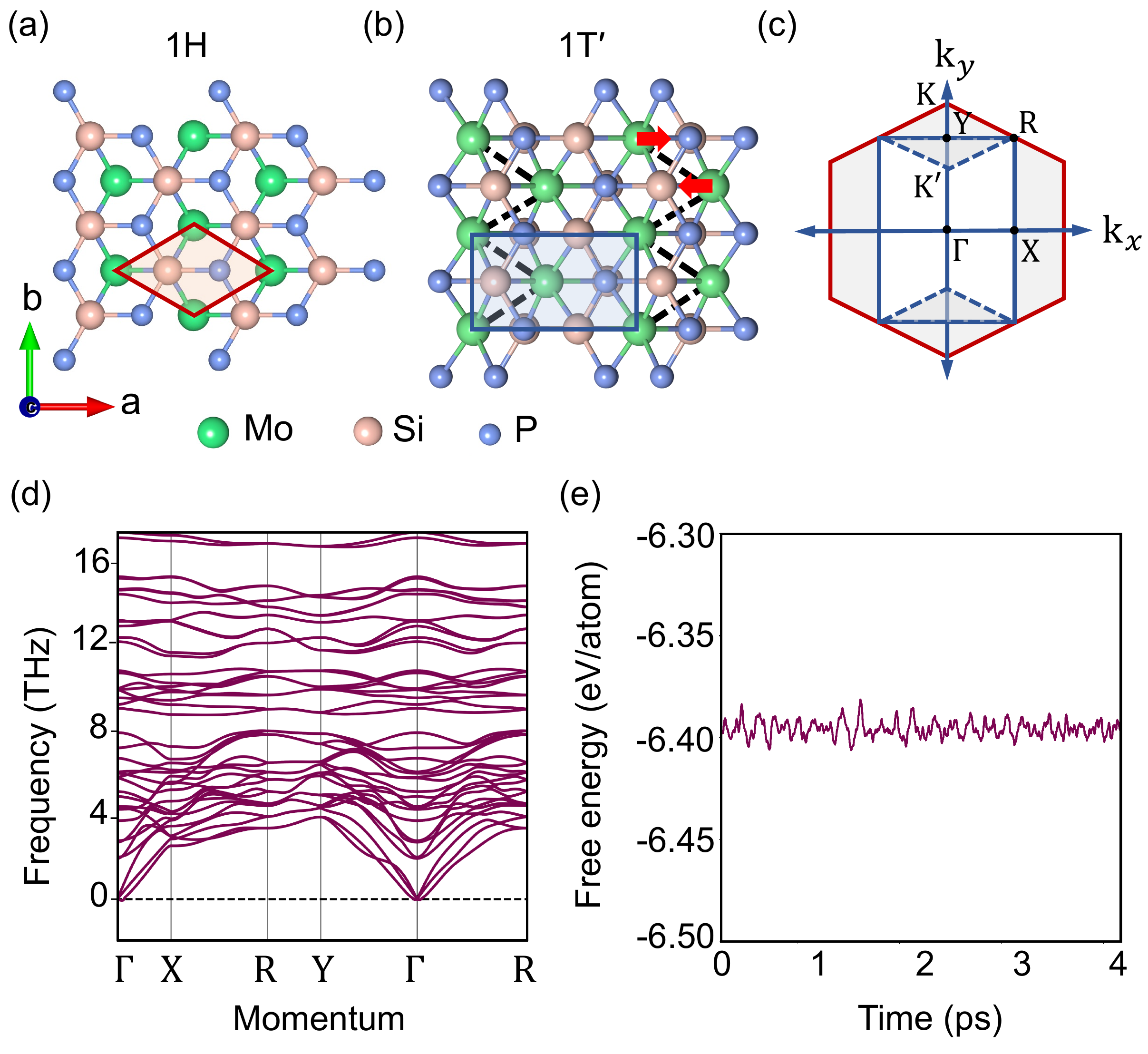}
 \caption{Crystal structure of MoSi$_2$P$_4$ monolayer in (a) 1H and (b) $1T^\prime$ phase. The unit cell of the 1H phase is indicated by the red rhombus in (a) and the $1T^\prime$ phase is shown with the blue rectangle in (b). The Mo atoms are distorted from their original hexagonal positions to form 1D zigzag chains along the $y$ axis in the $1T^\prime$ phase as shown with dashed black lines in (b). (c) The associated 2D Brillouin zones (BZs) with high-symmetry points. The $K$ points of the hexagonal BZ (red color) fold onto the $\Gamma-Y$ line of $1T^\prime$ rectangular BZ (blue color). (d) Phonon dispersion in 1T$^\prime$ MoSi$_2$P$_4$. (e) Total free energy of monolayer 1T$^\prime$ MoSi$_2$P$_4$ as a function of time step during the molecular dynamics simulation at $T= 300~K$.
 }\label{crystal_structure}
\end{figure}
The pristine phase of monolayer MSi$_2$Z$_4$ belongs to the 1H crystal structure family of 2D materials with space group D$_{3h}^1$ (P$\Bar{6}$m2, No. 187)~\cite{hong2020chemical,IslamMoSi2N4}. Figure \ref{crystal_structure}(a) shows its crystal lattice with taking MoSi$_2$P$_4$ as an exemplar system. The structure is layered along the hexagonal $c$ axis and can be viewed as a MoP$_2$ layer sandwiched between the two SiP layers, where the Mo atoms are located at the centre of the trigonal prism building block with six P atoms and the MoP$_2$ layer bonded vertically with the SiP layer~\cite{hong2020chemical}. In the 1T$^\prime$ phase of MoSi$_2$P$_4$ [Fig.~\ref{crystal_structure}(b)], the three atomic layers are locked in a way that the position of Mo atoms is at the centre of 60$^o$ twisted trigonal prism building block with six P atoms. This creates the octahedral local coordination of Mo atoms with the six P atoms in the MoP$_2$ layer but with different Mo-Mo bond lengths to form zig-zag atomic chains along the $y$ axis and period-doubling along the $x$ axis. Such a structural distortion lowers the hexagonal 1H symmetry to 1T$^\prime$ monoclinic symmetry with the space group P2$_1$/m~(No. 11) and forms a rectangular primitive unit cell as shown in Fig.~\ref{crystal_structure}(b). Importantly, the 1T$^\prime$ structure restores the inversion symmetry ${I}$ in contrast to the 1H phase. Figure~\ref{crystal_structure}(c) illustrates the BZs associated with both the 1H and 1T$^\prime$ phases where high-symmetry points are marked in both the pristine hexagonal and reduced rectangular BZs.

To determine the stability of polytypic structures, we present the calculated phonon dispersion of monolayer 1T$^\prime$ of MoSi$_2$P$_4$ in Fig.~\ref{crystal_structure}(d). The absence of imaginary frequency modes throughout the BZ indicates the dynamical stability of the 1T$^\prime$ phase. The structural stability is further substantiated by performing {\it ab-initio} molecular dynamics simulations at 300 K. Variation of the free energy as a function of the simulation time is presented in Fig.~\ref{crystal_structure}(e). The energy oscillates near a mean value of $~6.40$ eV/atom. However, the monolayer structure remains intact at the end of the simulations without any new reconstruction of the lattice, indicating the thermal stability of the monolayer. We have also checked the thermodynamical stability of other members of the 1T$^\prime$ MSi$_2$Z$_4$ family and found them to be stable, see Supplemental Material (SM). Similar to structural stability and experimental realization of the 1T$^\prime$ phase of TMDs, these results elucidate that the 1T$^\prime$-MSi$_2$Z$_4$ is stable and can be synthesized experimentally.

\begin{figure*}[ht!]
\centering
\includegraphics[width=0.8\textwidth]{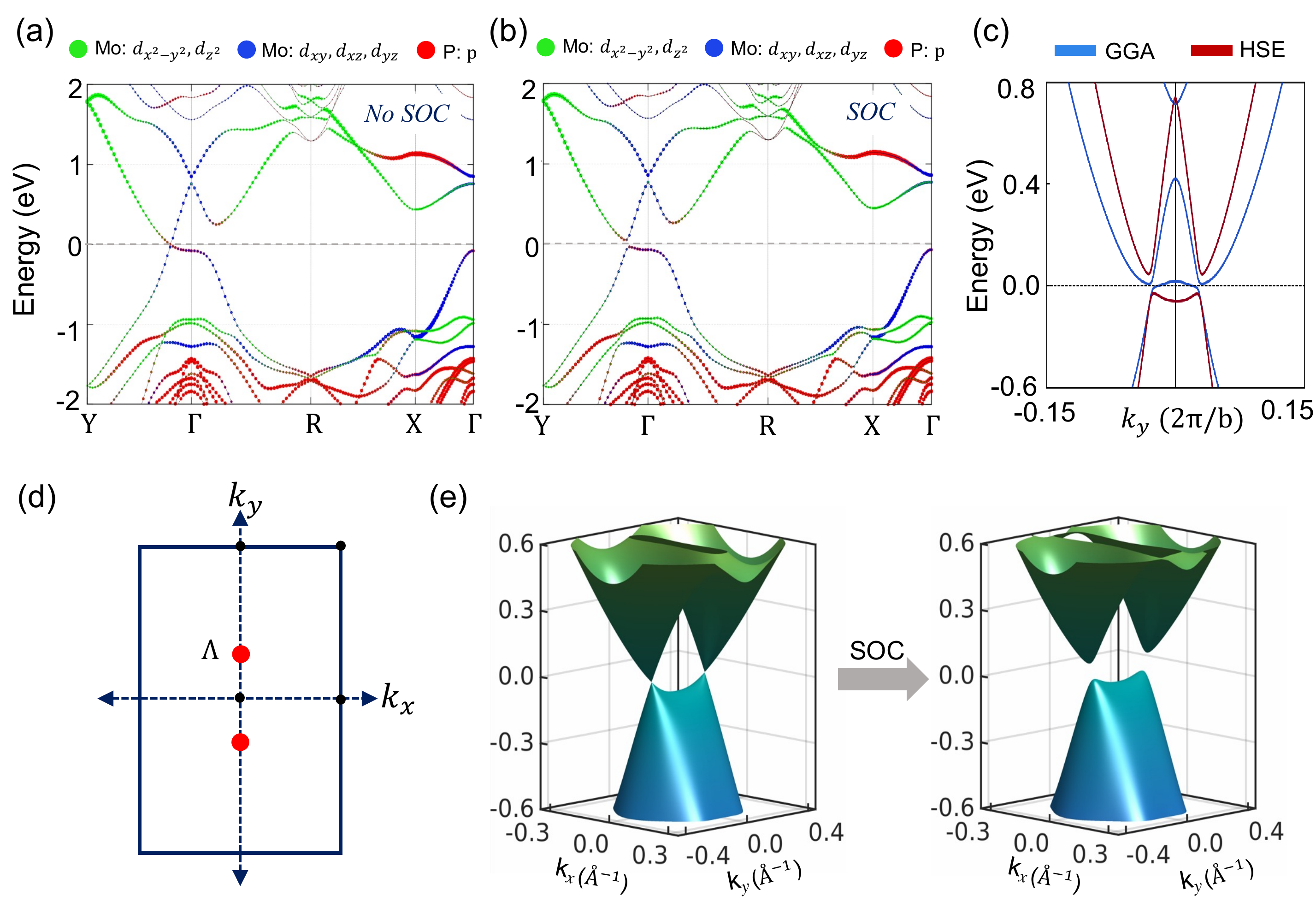}
 \caption{Band structure of 1T$^\prime-$MoSi$_2$P$_4$ (a) without and (b) with spin-orbit coupling using the HSE hybrid functional. The horizontal dashed line marks the Fermi level. The orbital compositions of bands are shown using different colors. (c) Closeup of bands obtained with HSE and GGA functionals along the $Y-\Gamma-Y$ line. (d) Location of valence and conduction band crossings at the Fermi level without spin-orbit coupling in the  2D Brillouin zone. (e) $E-k_x-k_y$ space rendition of spin-orbit 
-coupling-driven electronic structure crossover in monolayer 1T$^\prime-$MoSi$_2$P$_4$ with the HSE functional. }
\label{electronic_structure}
\end{figure*}

\begin{table*}
\caption{Calculated structural and electronic parameters of 2D 1T$^\prime$-MSi$_2$Z$_4$ (M = Mo or W and Z = P or As). The structural parameters include in-plane lattice constants $a$ and $b$, and interatomic separation $d_1$ and $d_2$ in the transition-metal zigzag chain. The electronic parameters present the global band gap obtained with GGA (E$_g^{GGA}$) and HSE (E$_g^{HSE}$), the inverted bandgap at the $\Gamma$ point calculated with GGA ($\delta^{GGA}$) and HSE ($\delta^{HSE}$). Summary of the topological state in 2D 1T$^\prime$-MSi$_2$Z$_4$ is also given. QSH denotes quantum spin Hall state. See text for more details.
}
\begin{tabular}{c c c c c c c c c c c}
\hline \hline 
 Material 		& a (\r{A})& b (\r{A})	&	$d_1$ (\r{A})	&	$d_2$ (\r{A}) 	& \multicolumn{2}{c}{Band gap (meV)} &  \multicolumn{2}{c}{Inverted Gap (meV)} & Topological  &Topological \\
 			&  			&    			 &    				&     			& 			   & 				& 				&     	 & invariant 		& state \\
			&  			&    			 &    				&     			& E$_g^{GGA}$  & E$_g^{HSE}$	& $\delta^{GGA}$  & $\delta^{HSE}$     &	 & 		 \\ 	\hline
MoSi$_2$P$_4$  	& 6.141	& 3.430        	&  2.945      &  4.194   & -10.4 		&  86.2   			& 398 		& 842		&  	$Z_2=1$ 	& QSH	\\
 MoSi$_2$As$_4$    	& 6.388	& 3.579       	& 3.012       &  4.362   & -39.5 		& 109.2      		& 391 		&800 		&  	$Z_2=1$ 	& QSH	\\

WSi$_2$P$_4$    	& 6.129	& 3.441       	& 2.939       & 4.133    &  -23.2 		& 198.5    			& 712 		& 1079		& 	 $Z_2=1$ 	& QSH	\\
WSi$_2$As$_4$	& 6.364  	& 3.589  		&  2.993      & 4.368     &  -0.07 		& 204.3     		& 675 		& 1058 		& 	$Z_2=1$	& QSH	\\
\hline  
\hline
\end{tabular} \label{T1:bulk}
\end{table*} 

We now discuss the orbital-resolved electronic structures computed with GGA and HSE hybrid functionals to delineate the topological state of various 1T$^\prime$-MSi$_2$Z$_4$ compounds. Figures~\ref{electronic_structure}(a)-(b) show the representative HSE band structure of 1T$^\prime$-MSi$_2$Z$_4$ taking MoSi$_2$P$_4$ as an example. The band structure is semimetallic with isolated spinless Dirac-type crossings on the $\Gamma-Y$ directions without SOC [Fig.~\ref{electronic_structure}(a)]. Adding relativistic effects to electronic structure opens a hybridization gap at these band crossings, thereby realizing a semiconducting state with a global bandgap E$_g^{HSE}$ of 86 meV. The valence and conduction band extremum points are located away from time-reversal invariant momentum (TRIM) points at $\Lambda = \pm 0.103${\AA}$^{-1}$ [red dots in ~\ref{electronic_structure}(c)] on the $\Gamma$-Y line, forming a camel-back-like band structure near the $\Gamma$ point~\cite{PhysRevResearch.4.023114}. Such band structures generally imply a nontrivial topology. We find that the $p$ states of P lie below the Mo $d$ states with a clear band inversion at the $\Gamma$ point, strong hybridization between the P and Mo states notwithstanding. This unusual orbital ordering is driven by a structural transition from 1H to 1T$^\prime$, which lowers the energy of the transition-metal states and results in a large inverted bandgap $\delta^{HSE}$ of 842 meV at the $\Gamma$ point that is larger than the existing 1T$^\prime$ QSH materials. The calculated electronic and structural parameters of our investigated 1T$^\prime$-MSi$_2$Z$_4$ materials are listed in Table ~\ref{T1:bulk} with band structures shown in the SM. Since monolayer 1T$^\prime$-MSi$_2$Z$_4$ respects inversion symmetry, we calculated the $Z_2$ invariant from the parity eigenvalues of the occupied states at the TRIM points and found $Z_2=1$ (nontrivial) for all investigated materials. These results indicate that 2D 1T$^\prime$-MSi$_2$Z$_4$ monolayers are QSH insulators. 

\begin{figure}
\includegraphics[width=0.49\textwidth]{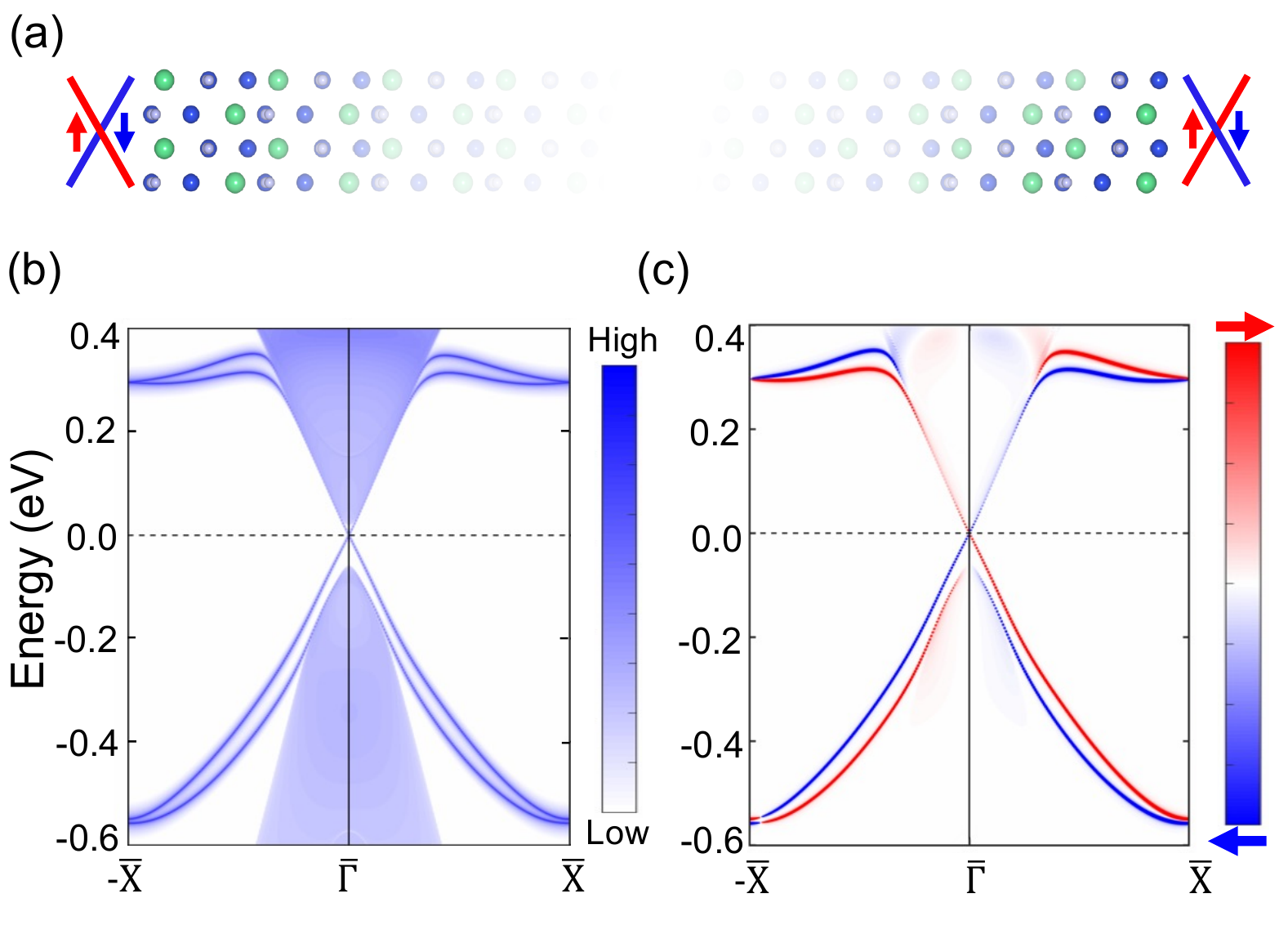}
 \caption{(a) Lattice structure of semi-infinite one-dimensional edges of 1T$^\prime$-MoSi$_2$P$_4$. Spin-polarized edge states are shown schematically on the left and right edges. (b) Electronic spectrum of (010) edge. Nontrivial edge states can be seen in the 2D bulk bandgap. (c) Calculated left-edge state spin-texture of 2D 1T$^\prime$-MoSi$_2$P$_4$. Red and blue colors indicate the up and down spin polarization, respectively. }
\label{surface_state}
\end{figure}

We present the band structures obtained using HSE06 (red curves) and GGA (blue curves) along the Y-$\Gamma$-Y directions in Fig.~\ref{electronic_structure}(b) to estimate the bandgap corrections in 1T$^\prime$-MSi$_2$Z$_4$ monolayers. The hybrid functionals are generally considered more accurate in estimating band bending, band order, and bandgap in comparison to the GGA functional. We find that while HSE06 corrects the inverted bandgap at the $\Gamma$ point in comparison to GGA, the overall band structures obtained with the two functionals are topologically equivalent with a band inversion at the $\Gamma$ point. Figure~\ref{electronic_structure}(d) presents the formation of QSH state in 1T$^\prime$-MSi$_2$Z$_4$ monolayers by switching off the SOC. Specifically, the gapless band crossings are found at finite momenta along the $Y-\Gamma-Y$ line at the  $\Lambda$ points. Switching on the SOC, hybridizes these band crossings to generate the QSH state. These results imply that the band inversion in 2D 1T$^\prime$-MSi$_2$Z$_4$ emerges via the structural transition while the SOC is responsible for forming the QSH state.

The spin-polarized edge states protected by the time-reversal symmetry are the hallmark of the QSH state. To highlight these states, we plot the calculated edge spectrum and the associated spin-texture of MoSi$_2$P$_4$ in Fig.~\ref{surface_state}. Figure~\ref{surface_state}(a) depicts the terminated left and right edges along the $y$ axis with schematics of the nontrivial states of our $xy$ monolayer ($x$ remains the periodic direction). Since the two edges are related by inversion symmetry, we display states for the left edge in Figs.~\ref{surface_state}(b)-(c). A pair of counter-propagating states with opposite spin-polarization are seen inside the bandgap with a Dirac point at $\Gamma$. Similar results are found for other 1T$^\prime$-MSi$_2$Z$_4$ materials. 

\begin{figure*}[ht!]
\includegraphics[width=0.8\textwidth]{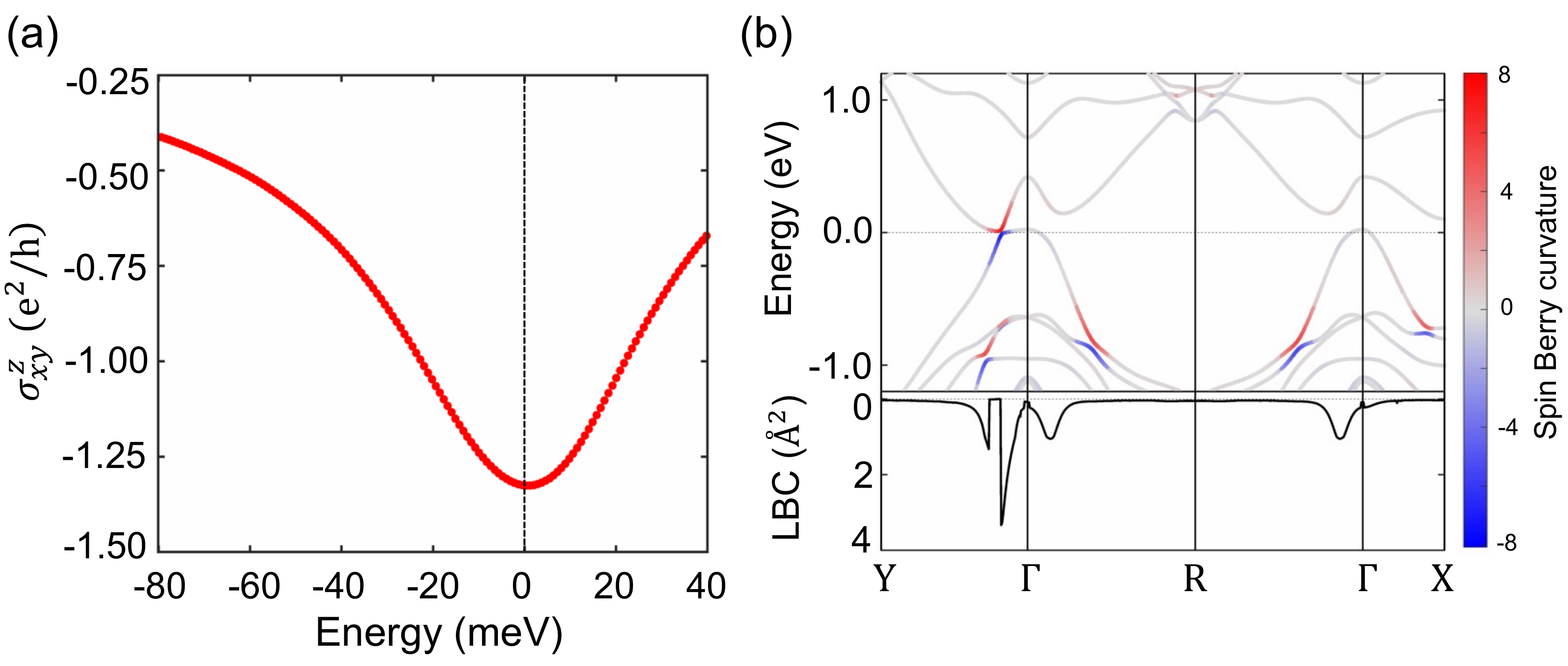}
\caption{(a) Intrinsic spin Hall conductivity (SHC) $\sigma_{xy}^z$ as a function of Fermi energy for monolayer 1T$^\prime$-MoSi$_2$P$_4$. SHC is given in units of $\frac{e^2}{h}$. Value of the SHC is maximum at the Fermi level. (b) Band-resolved (top) and $k-$resolved (bottom) spin Berry curvature of monolayer MoSi$_2$P$_4$ along the high-symmetry directions in the 2D Brillouin zone. LBC stands for logarithm of $\Omega_{xy}^z(\bm{k})$ (Eq.~\ref{ksbc}) and the colorbar refers to the logarithm of $\Omega_{n,xy}^z(\bm{k})$ (Eq.~\ref{SBC}).
} \label{Spin_Hall_Conductivity}
\end{figure*}

Having established the QSH state in 1T$^\prime$-MSi$_2$Z$_4$ monolayers, we now discuss their intrinsic spin Hall conductivity (SHC). We obtain the SHC $\sigma_{xy}^{z}$ using the Kubo formula~\cite{SHC_3d2d1d, shc_TaAs, ahc_weyl_Jianwei, shc_weyl_tmd,SHC_w90} as:
\begin{equation} \label{shc}
\sigma_{xy}^{z}=\frac{-e^2}{\hbar} \frac{1}{A} \sum_k \Omega_{xy}^z (\bm{k})
\end{equation} 
where,
\begin{equation}\label{ksbc}
\Omega_{xy}^z(\bm{k})= \sum_n f_{n} (\bm{k})\Omega_{n,xy}^z(\bm{k})
\end{equation}
is the $k$-resolved spin Berry curvature and
\begin{equation}\label{SBC}
\Omega_{n,xy}^z(\bm{k})= {\hbar^2} \sum_{m\ne n} 
\frac {-2\text{Im}\langle n\bm{k}|  \hat{J_x^z} |m\bm{k}\rangle \langle m\bm{k}|  \hat v_y |n\bm{k}\rangle}
{(E_{nk}-E_{mk})^2}
\end{equation}
is the band resolved spin Berry curvature.

In Eqs.~\ref{shc}-\ref{SBC},  $A$ is the area of 2D unit cell and $|n\bm{k}\rangle$ denotes the Bloch state with energy $E_{nk}$ and occupation $f_{n}(\bm{k})$. The spin current operator $\hat{J_x^z}=\frac{1}{2}\{\hat\sigma_z, \hat v_x\}$ with the spin operator $\hat\sigma_z$ and the velocity operator $\hat v_x$. The SHC $\sigma_{x,y}^{z}$ represents the spin-current along the $x$ direction generated by the electric field along the $y$ direction, and the spin current is polarized along the $z$ direction. We consider a dense grid of $10^{6}$ $k-$points in conjunction with maximally-localized Wannier functions to evaluate the spin Berry curvature and SHC. Figure~\ref{Spin_Hall_Conductivity}(a) presents the calculated SHC as a function of the Fermi energy. The SHC is maximum near the band crossing points [marked with dashed line] reaching the highest value of $\sim1.3\frac{e^2}{h}$. The amplitude of SHC decreases quickly away from the band crossings points. This can be further seen from our band and $k-$resolved spin Berry curvature in Fig.~\ref{Spin_Hall_Conductivity}(b). The spin Berry curvature is largely concentrated near the valence and conduction band-crossing point along the $\Gamma-Y$ direction having a SOC-driven hybridization gap. Notably, the perfect quantization of the SHC cannot be expected in QSH materials since they violate the $S_z$ spin conservation \cite{SHC_quantization}. However, MoSi$_2$P$_4$ shows a larger value of SHC than the other 1T$^\prime$ TMDs with the QSH state. An optimal setup to exploit the large SHC in MoSi$_2$P$_4$ should be in the clean limit with the Fermi level lying between the band crossings points.     

\begin{figure}[b!]
\centering
\includegraphics[width=0.49\textwidth]{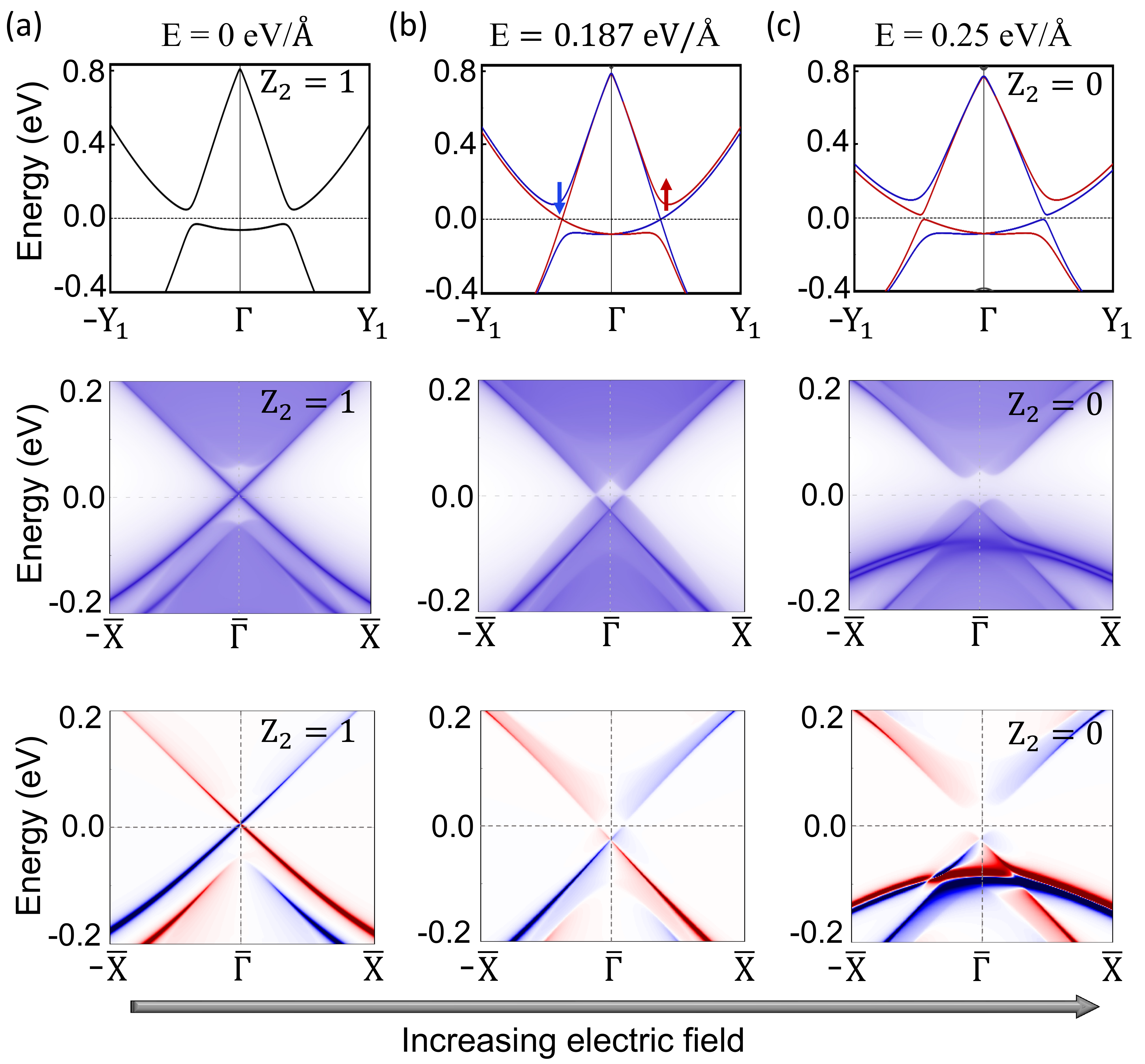}
\caption{Band structure of 1T$^\prime-$MoSi$_2$P$_4$ monolayer for various values of the vertical electric field $E_z$: (a) 0, (b) 0.187, and (c) 0.250 eV/\AA. The top, middle, and bottom rows show 2D band structure, (010) edge spectrum, and the edge-state spin-texture, respectively. Red and blue colors identify up and down spin states respectively. }
\label{OUTPLANE_electricfield}
\end{figure}

\begin{figure}[t!]
\centering
\includegraphics[width=0.49\textwidth]{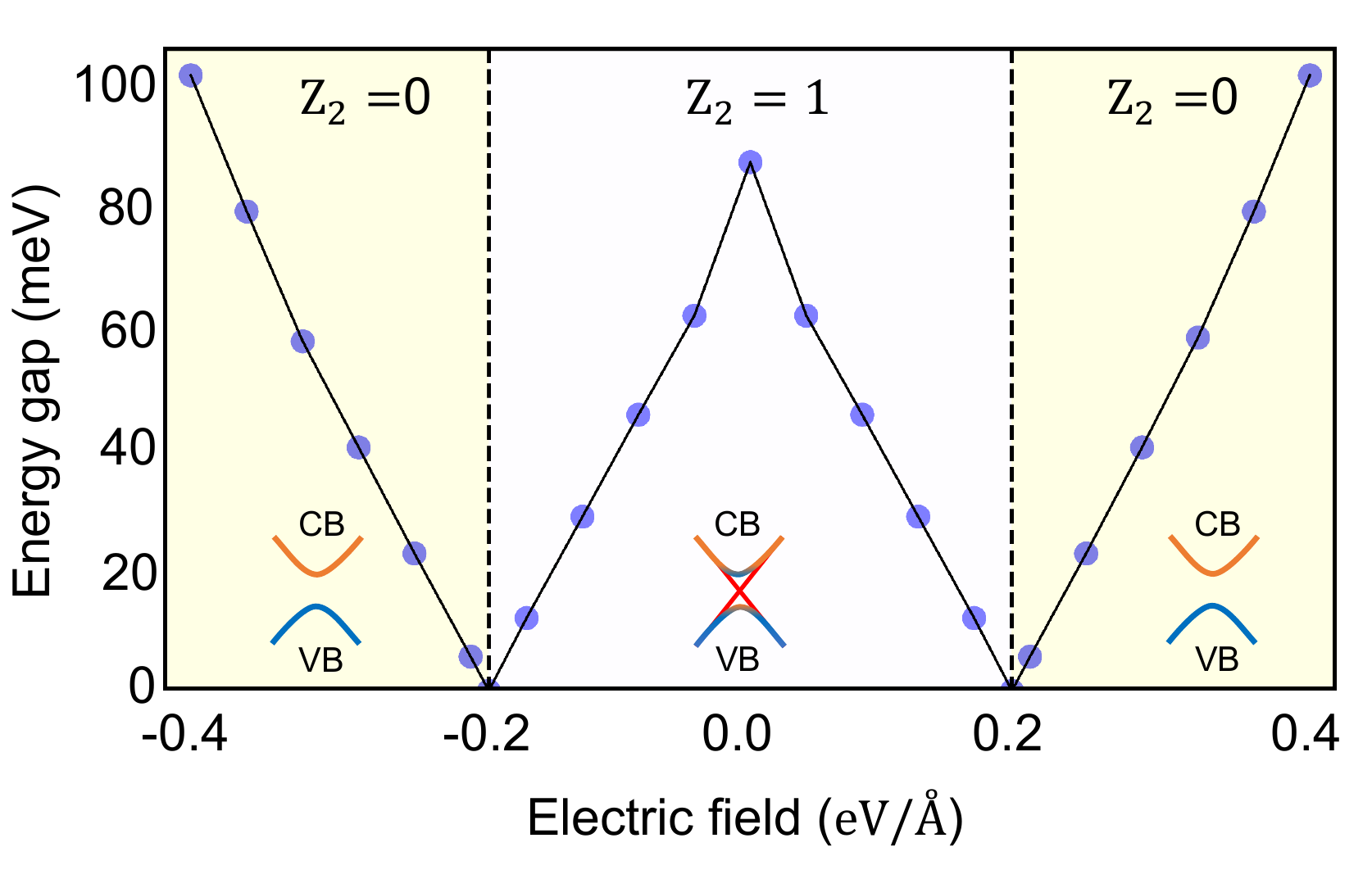}
\caption{Topological phase diagram of 1T$^\prime-$MoSi$_2$P$_4$ as a function of the vertical electric field ($E_z$). Critical electric field strength for the topological phase transition from $Z_2= 1$ to $Z_2= 0$ state is marked with vertical dashed lines. }
\label{gapwithelectricfield}
\end{figure}

We now demonstrate the tunability of the QSH state and the switching of the topological state under a vertical electric field. The topologically inverted bands between the transition metal $d$ and pnictogen's $p$ orbitals lie at well-separated 2D planes in 1T$^\prime$ monolayer. This distinct spatial location of bands provides a natural basis for their tunability by an out-of-plane (vertical) electric field $E_z$. Figure~\ref{OUTPLANE_electricfield} shows the HSE band structure for various electric field values. The electric field induces Rasbha spin-splitting in the states by breaking inversion equivalence on the top and bottom sides of the monolayers. This is evident from spin-split states shown with distinct colors in the top row of Figs.~\ref{OUTPLANE_electricfield}(b)-(c). As the electric field increases, the band gap decreases to zero at the critical electric field value of E$_c$ = 0.187 eV/{\AA} where the spin-up and spin-down bands cross at the opposite $\Lambda$ points. With further increase in the electric field, the bandgap reopens. An analysis based on the $Z_2$ invariants and edge-state dispersions (Fig.~\ref{OUTPLANE_electricfield}) shows that this bandgap closing drives a change in the topology to a trivial state with $Z_2=0$. This topological phase transition destroys the topological edge states thereby switching the QSH state in 1T$^\prime-$MoSi$_2$P$_4$. A change in polarity of the electric field shows a similar topological phase transition to a trivial state. The evolution of QSH and trivial insulator state as a function of the applied electric field is displayed in Fig.~\ref{gapwithelectricfield}.

The aforementioned results indicate an electric field on/off control of the spin-polarized edge currents in 1T$^\prime-$MSi$_2$Z$_4$ in similar to the case of 1T$^\prime-$TMDs~\cite{LiangFu}. Since the crystal symmetries of both these materials families are the same, various device ideas conceived for 1T$^\prime$-TMDs can be applied to the MSi$_2$Z$_4$ family with the added advantage of a large inverted bandgap and large SHC. For example, monolayer MSi$_2$Z$_4$ could be interfaced with a large bandgap 2D insulator to protect the helical edge channels from being gapped by interlayer hybridization to realize a topological transistor~\cite{LiangFu}. When the Fermi level is placed in the nontrivial bandgap, a nearly quantized SHC would be realized in this device under zero or small electric fields. An electric field beyond the critical value of $\pm~0.187$ eV/{\AA} can switch off the quantized spin-Hall conductance, driving it into a trivial insulating state.  

{\it Summary:--}
We have demonstrated the existence of a tunable QSH state with a large bandgap in a new polytypic structure of the recently introduced bottom-up synthesized MSi$_2$Z$_4$ family of 2D materials. Our analysis based on phonon spectra and molecular dynamics simulations shows that these materials realize a thermodynamical stable 1T$^\prime$ phase in addition to the putative 1H phase. Our in-depth electronic structure modeling reveals that a structural distortion in the 1H phase leads to the 1T$^\prime$ structure and induces a topological band inversion. An  inverted bandgap as large as $204$ meV is found in the MSi$_2$Z$_4$ family that is even larger than in the existing 1T$^\prime-$TMDs hosting a QSH ground state. Our calculated SHC shows a large value of $\sim1.3\frac{e^2}{h}$ in MoSi$_2$P$_4$ that arises from the large spin Berry curvature induced by spin-orbit-split bands at the band inversion points. We also show that the QSH state is tunable with a vertical electric field, which provides an external control for switching or turning on/off the QSH state. Our study thus not only introduces a new polytypic structure of recently introduced 2D MSi$_2$Z$_4$ materials, which supports a large bandgap QSH state, but it also shows that this new materials family will provide an excellent platform for realizing nontrivial states with large spin Hall conductance.

{\it Acknowledgements:--}
This work is supported by the Department of Atomic Energy of the Government of India under project number 12-R$\&$D-TFR-5.10-0100. The work at Institute of Physics, Polish Academy of Sciences is supported by the Foundation for Polish Science through the International Research Agendas program co-financed by the European Union within the Smart Growth Operational Programme and the National Science Center in the framework of the "PRELUDIUM" (Decision No.: DEC-2020/37/N/ST3/02338). We acknowledge the access to the computing facilities of the Interdisciplinary Center of Modeling at the University of Warsaw, Grant G84-0, GB84-1 and GB84-7. We acknowledge the CINECA award under the ISCRA initiative IsC93 "RATIO" and IsC99 "SILENTS" grant, for the availability of high-performance computing resources and support. The work at Northeastern University was supported by the Air Force Office of Scientific Research under award number FA9550-20-1-0322, and benefited from the computational resources of Northeastern University's Advanced Scientific Computation Center (ASCC) and the Discovery Cluster.

\bibliography{MoSi2N4_BS.bib}

\end{document}